\documentclass[12pt]{article}
\usepackage{amsmath,cite,epsf,graphicx}

\def\beq{\begin{equation}}
\def\eeq{\end{equation}}
\def\bea{\begin{eqnarray}}
\def\eea{\end{eqnarray}}

\input{babarsym}

\newcommand{\mx} {\ensuremath{m_{X}}\xspace}
\newcommand{\mxqsq} {\ensuremath{(m_{X}, q^2)}\xspace}
\newcommand {\pplus}  {\ensuremath{P_{+}}\xspace}
\newcommand{\elsmax}{\ensuremath{(E_{\ell},s_{\mathrm{h}}^{\mathrm{max}})}}
\newcommand{\smax}{\ensuremath{s_{\mathrm{h}}^{\mathrm{max}}}}
\newcommand{\el} {\ensuremath{E_{\ell}}\xspace}
\newcommand {\mb}{\ensuremath{m_b}\xspace}
\newcommand {\mc}{\ensuremath{m_c}\xspace}
\newcommand{\btoxlnu}{\ensuremath{\B \to X_u {\ell} \nu_{\ell}}}
\newcommand{\btoxclnu}{\ensuremath{\B\to X_c \ell \nu_{\ell}}}

\begin{document}

\begin{titlepage}
\begin{flushright}
{\small
ROME1/1461-07
\\
DSNA/33-2007
\\
%0711.0860 [hep-ph]
}
%\\
\end{flushright}
\vskip 1cm
\begin{center}
{\Large  \bf \boldmath
Inclusive Measure of \Vub\ with the\\[2mm] Analytic Coupling Model
 } \\[20mm]
 \bf     U.~Aglietti$\,^a$, ~ F.~Di Lodovico$\,^b$, ~ G.~Ferrera$\,^c$, ~
G.~Ricciardi$\,^d$
\\ [2mm]
\vskip 0.5cm
{$^a$ \it  Dip.\ Fis., Univ.\ di Roma I ``La Sapienza'' \& INFN Roma,
 Roma, Italy
\\[2mm]
$^b$  Queen Mary, University of London,  Dep. of Phys., London, UK
\\[2mm]
$^c$  Dip. Fis., Univ.\ di Firenze \& INFN Firenze, Sesto Fiorentino, Firenze,
Italy
\\[2mm]
$^d$ Dip.\ Scienze Fis., Univ.\ di Napoli ``Federico II'' \& INFN Napoli,
Napoli, Italy
}
\\
\end{center}
\vskip 1.5cm
\hrule
\begin{abstract}
\vspace{.1cm} By analyzing \btoxlnu\ spectra with a model based on
soft--gluon resummation and an analytic time--like QCD coupling,
we obtain \beq \Vub \, = \, ( \, 3.76 \, \pm \, 0.13 \, \pm \,
0.22 \, ) \times 10^{-3}, \nonumber \eeq where the first and the
second error refers to experimental and theoretical errors,
respectively.\ This model successfully describes the accurate
experimental data in beauty fragmentation, which has similar
soft-gluon effects. \ The \Vub\ value is obtained from the
available measured semileptonic branching fractions in limited
regions of the phase--space. The distributions in the lepton
energy \el, the hadron invariant mass \mx, the light--cone
momentum $\pplus \!\equiv E_X - |\vec{p}_X|$, together with the
double distributions in \mxqsq\ and \elsmax, are used to select
the phase--space regions. The $q^2$ is the dilepton squared
momentum and \smax\ is the maximal $\mx^2$ at fixed $q^2$ and \el
. The \Vub\ value obtained is in complete agreement with the value
coming from exclusive $B$ decays and from an over--all fit to the
Standard Model parameters. We show that the slight disagreement
(up to $+2\,\sigma$) with respect to previous inclusive measurements is
not related to different choices for the $b$ (and $c$) masses, but
to a different modelling of the threshold (Sudakov) region.

\end{abstract}
\hrule
\vspace{-10mm}
%\noindent {
%%\footnotesize
%\tiny PACS: }

\end{titlepage}

\setcounter{footnote}{0} \setcounter{page}{2} \setcounter{section}{0}
\newpage

\section{Introduction}

By comparing various spectra in the decays \beq \label{start} \B
\, \to \, X_u \, + \, \ell \, + \, \nu_{\ell} \eeq with the
predictions of a model including non--perturbative corrections to
soft--gluon dynamics through an effective QCD
coupling~\cite{model}, we obtain a value for the $V_{ub}$
Cabibbo--Kobayashi--Maskawa (CKM) matrix
element~\cite{ckm,grinstein} \beq \Vub \, = \, ( \, 3.76 \, \pm \,
0.13 \, \pm \, 0.22 \, ) \times 10^{-3} \, , \eeq where the first
and the second error refers to experimental and theoretical
errors, respectively. The model basically involves the insertion,
inside standard threshold resummation formulae, of an effective
QCD coupling $\tilde{\alpha}_S(k^2)$; such coupling is based on an
analyticity requirement and includes resumming absorptive effects
in gluon cascades~\cite{shirkov}. A significative point is that
the model, which has no free unknown parameters, describes rather
well also \B--meson fragmentation data at the $Z^0$ peak
~\cite{acf}, where
--- unlike $B$ decays ---
accurate data are available and there is no uncertainty coming
from CKM matrix elements. The main properties of the model are
sketched in sec.~\ref{Threshold--Resummation}.

We analyze the distributions in the lepton energy \el, the final
hadron invariant mass \mx, the light--cone momentum $\pplus \equiv
E_X - |\vec{p}_X|$, together with the double distributions in
\mxqsq\ and in \elsmax, with $q^2$ being the dilepton
squared momentum and \smax\ the maximal $\mx^2$ at fixed $q^2$ and \el.

The decay rates for the quark--level transitions
$ \b \to u\, \ell\, \nu_\ell $ are proportional to $\Vub^2$ and
 two different methods are selected for measuring this matrix element.
In the first method, one considers a specific exclusive decay
%, such
%as for example \beq \label{exclu} \B \, \to \, \rho \, \ell \, \nu_\ell
%\, ,  \eeq
by identifying experimentally the final--state hadrons. Dynamics
is substantially non perturbative and current theoretical
predictions use QCD sum rules, quark models, lattice QCD, etc. The
second method involves the inclusive hadron final states $X_u$ in
eq.~(\ref{start}). In general, given a kinematical variable $p$,
such as for example the energy \el\ of the charged lepton, one
measures the number of \B's decaying semileptonically to $X_u$
with $p$ in some interval $(a,b)$, divided by the total number of
produced \B's (decaying into any possible final state): \beq
\label{br} {\cal B}\left[ p \in (a,b) \right] \, \equiv \, \frac{
N [ B \, \to \, X_u \, \ell \, \nu_{\ell} , ~ p \in (a,b) ] } {
N\left[ B \to {\rm (anything)} \right] } \, . \eeq Since the
beauty mass $m_b \approx 5$~\gev\ is rather large compared to the
hadronic scale, %\beq
$m_b \gg \Lambda $,
%\eeq
one can consider semileptonic \b--decays as hard processes, to be
treated in perturbative QCD, with inclusive hadron final states
coming from gluon radiation. Because $ |V_{cb}|^2/|V_{ub}|^2 \,
\approx \, 10^{\, 2} \, , $ $b \, \to \, c \, \ell \, \nu_{\ell}$
decays constitute a huge background to $b \, \to \, u \, \ell \,
\nu_{\ell}$ ones as far as inclusive quantities are
concerned\footnote{ The non--vanishing charm mass reduces the $b
\, \to \, c \, \ell \, \nu_{\ell}$ rate roughly by a factor two.
}. To avoid (or at least substantially reduce) such background,
one has to consider kinematical regions where $b \to c$
transitions are kinematically forbidden (or at least strongly
disfavored):
typically %, one has to consider
end--point regions.
% for the variable
%$p$.
On the theoretical side, this restriction {\it has a price},
because the available phase--space to QCD partons gets strongly
reduced. One ends up with the so--called threshold region
\cite{Altarelli:1982kh},
defined as having parametrically\footnote{ This region is also called
Sudakov region, large--$x$ region and radiation--inhibited region.
} \beq m_X \, \ll \, E_X \, . \eeq The perturbative expansion of
spectra in the threshold region is affected by large logarithms
$\approx \alpha_S^n \log^{2n} (2 E_X/m_X$), which must be resummed
to all orders in $\alpha_S$ in order to have a reliable result
\cite{mp,ugogiu2004,noi}. Consistent inclusion of subleading
logarithms requires a prescription for the QCD coupling in the
low--energy region $\sim \Lambda$ --- in principle completely
arbitrary
--- which in our model is the analyticity condition. Furthermore,
Fermi motion, a genuine non--perturbative effect related to a
small vibration of the $b$ quark in the $B$ meson, comes into play
when $m_X$ becomes as small as $\approx \sqrt{\Lambda \, E_X\,}$.

%Let us compare virtues and short--comings of the two methods for
%measuring $|V_{ub}|$.
The exclusive determination uses a smaller
event sample because it deals with a single channel, with the
consequence that larger statistical errors are expected. Since the
relevant hadronic matrix elements %for eq. (\ref{exclu}), $\langle \rho
%| J_{\mu}^{b \to u} | B \rangle$,
can be computed in this case with a first--principle technique,
namely lattice QCD, one expects that, with increasing computing
resources, hadronic uncertainties can be systematically (and
almost arbitrarily) reduced. On the contrary, the inclusive method
suffers less from statistics, but needs a modelling of
non--perturbative QCD effects, which cannot be completely derived
from first principles.
%in explicit form. To summarize,
Asymptotically in time, we expect the exclusive measure to take
over the inclusive one.

At present, the determinations from inclusive and exclusive decays are
given with a relative precision of about 5--8\% and 16\%~\cite{hfag},
respectively, using different computations for the inclusive decays and an
average of the Lattice QCD determinations for the exclusive decays giving
\beq
|V_{ub}| = (3.51\pm 0.21^{+66}_{-42}) \times  10^{-3}.
\eeq
There is a $\approx +1-2\,\sigma$ discrepancy between the inclusive
and exclusive measurements, depending on the calculation used for
the inclusive decays, where the value of $|V_{ub}|$ obtained by the exclusive calculation
is always larger than the corresponding measurement obtained by the
exclusive decays, indicating some ``tension'' between the above methods.

A third independent measure of \Vub\ stems from a general fit
of the Standard Model (SM) parameters.
One assumes the validity of the Standard Model
--- and therefore also the unitarity of the CKM matrix ---
without using the direct inclusive or exclusive determinations.
The result is~\cite{utfit}: \bea |V_{ub}| &=& ( 3.44 \pm 0.16 )
\times 10^{-3} \quad \mbox{(SM fit)} \, . \eea The global fit of
the SM therefore ``prefers'' the exclusive
determination, % ($0.14\,\sigma$),
while the inclusive one is in agreement at
$\approx 3\,\sigma$ level only.

It has been suggested that the discrepancy between the value of
the experimental measurement and the inclusive theoretical
prediction could signal effects of new physics from extra Higgs
particles~\cite{nonSM}. In our opinion, the above discrepancy does
not necessarily imply a signal of new physics. In other words, we
believe that the ``tension'' can be dynamically explained inside
the Standard Model.
Even though there are several models in
literature describing non--perturbative effects in inclusive $B$
decays, which give results perfectly consistent with each other
\cite{rothstein,lange, ligeti,neubert,gardi,gambino}, we believe that
a possible interpretation of the above scenario is that the theoretical
uncertainties have been under--estimated.
A re--analysis of the same data with a rather different model
may therefore be useful.
To re--extract \Vub\ in this spirit, it is convenient to
identify the different dynamical effects which come from theory
and cannot be extracted from the data. As it will be explicitly
shown in sec.~\ref{meth},  one has to compute, roughly speaking,
both:
\begin{enumerate}
\item inclusive rates, (strongly) dependent on the choice of the
heavy quark masses $m_b$ (and $m_c$),
as well as on the QCD coupling at a reference scale
(typically $\alpha_S(m_Z)$).
In sec.~\ref{meth} we present two methods which differ in
the treatment of the inclusive quantities; we also discuss our
choices of the $b$ and $c$ masses; \item suppression factors, for
the restriction of the kinematical variable $p$ in some
experimentally accessible range. These factors are affected by
large threshold logarithms and by the related Fermi--motion
effects mentioned earlier.
\end{enumerate}
The discrepancy of our analysis with respect to previous ones does
not rely on the estimate of inclusive quantities (different
choices of quark masses, of $\alpha_S(m_Z)$, etc.), i.e. on point
1., but on the modelling of the threshold region, i.e. on point 2.

In Sec. 3.1 we discuss our choices of quark masses  for describing
point 1. effects. The method used for the point 2. suppression
factor is discussed in ref.~\cite{model}. Here we just mention
that within this model the standard perturbative QCD resumming
description is modified in a minimal way in order to include some
infrared non perturbative effects. Since in this contest there is
no scale separation between perturbative and non perturbative QCD,
the model is only sensitive to the B meson mass.

Sec.~\ref{res} contains our results for \Vub\ coming from the
various distributions considered together with a discussion, while
sec.~\ref{concl} presents our conclusions.

\section{\!Threshold resummation with an effective coupling}
\label{Threshold--Resummation}

Let us briefly describe in this section the phenomenological model used
to extract $|V_{ub}|$, namely
threshold resummation with a {\it time-like} QCD coupling for the
semi--inclusive $B$ decays given in eq.~(\ref{start}) (for a more detailed discussion we refer
the reader to ref.~\cite{model}).

The first step is the construction of a general analytic
QCD coupling from the standard one, by means of an
analiticity requirement.
By requiring that the analytic coupling has the same discontinuity
of the standard coupling and no other singularity, one obtains
for example at one loop:
\begin{equation}
\label{basic}
\bar{\alpha}_S(Q^2) \, = \, \frac{1}{\beta_0}
\left[
\frac{1}{\log Q^2/\Lambda^2} - \frac{\Lambda^2}{Q^2 - \Lambda^2}
\right]
\end{equation}
The coupling above has no Landau pole, which has been subtracted
by a power correction and it is also immediate to check that it
has the same discontinuity of the standard one for $Q^2 < 0$,
i.e. in the time-like region, related to gluon branching.
The last term on the r.h.s. of eq.~(\ref{basic})
produces a series of power corrections once it is expanded for
$Q^2 \gg \Lambda^2$.
%A model for non perturbative effects can be constructed
%just by inserting the effective coupling (\ref{basic}) in the
%standard threshold resummation formula.
%: that has actually been made
%in \cite{Karanikas:2001cs} to describe the pion form factor.
It is then clear that using the effective coupling (\ref{basic})
in the standard threshold resummation formula,
power corrections to the QCD form factor
originate from the power corrections in the
effective coupling through integrations over transverse ($k^2$) and
longitudinal ($y$) degrees of freedom.
%Let us also note that the Mellin transform $y \to N$ in eq.~(\ref{sf})
%\footnote{
%That is to be contrasted with the standard logarithmic expansion,
%in which the Mellin transform can be computed in approximate way
%with a $\theta$-function \cite{mp}.
%}
%and the inverse Mellin transform $N \to y$ have to be computed
%exactly (in numerical way), in order to keep the effects of the power
%corrections.
%However, we have found that fragmentation data are better
%described by a specific variant of the above model, which
%incorporates an additional physical effect. The effective coupling
%(\ref{basic}) is supposed to model the evolution of a time-like
%gluon (emitted from a primary quark) into a jet.
Since in semi-inclusive decays the gluon
is always time-like, we have also included the absorptive parts of the
gluon polarization function (the well-known ``$-i\pi$'' terms)
into the effective coupling: that amounts to a resummation of
constant terms to all orders. At one-loop for example one obtains:
\bea \label{tl} \tilde{\alpha}_S(k^2) &=& \frac{1}{\beta_0} \left[
\frac{1}{2} \, - \, \frac{1}{\pi} {\rm arctan} \left( \frac{\log
k^2/\Lambda^2}{\pi} \right) \right] \, =
\nonumber\\
&=& \frac{1}{\beta_0 \log ( k^2/\Lambda^2 ) } \, - \,
\frac{\pi^2}{3}  \, \frac{1}{\beta_0 \log^3 ( k^2/\Lambda^2 ) } \,
+ \, {\mathcal O}\left[ \frac{1}{\beta_0 \log^5 ( k^2/\Lambda^2 )
} \right] \, ,
\eea
\begin{figure}[ht]
\begin{center}
\includegraphics[width=0.5\textwidth]{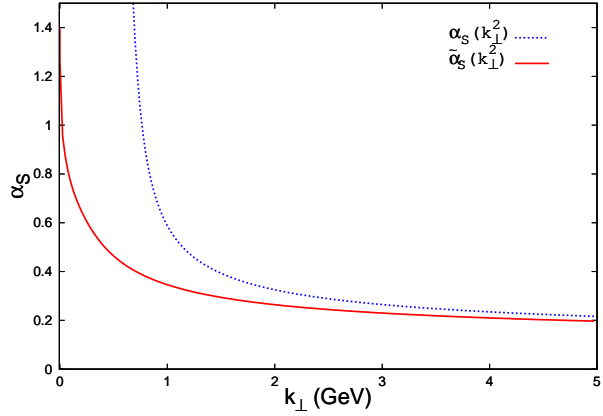}
\footnotesize\caption{
\label{fig:coup}
\it QCD couplings in NNLO. Dotted line (blue):
standard coupling $\alpha_S(k_{\perp}^2)$;
%dotted line (blue): analytic space-like coupling $\bar{\alpha}_S(Q^2)$;
continuous line (red):
analytic time-like coupling $\tilde{\alpha}_S(k_{\perp}^2)$.}
\end{center}
\end{figure}
%\clearpage
%It is remarkable that the inclusion of the $``-i\pi$'''s totally hides
%the power corrections originally present in the coupling (\ref{basic}):
%the effective coupling (\ref{tl}) has inverse powers of the logarithm
%to any order, but not even a single power corrections.
%Let us remark, however, that the time-like coupling cannot be
%consistently derived directly from the standard one, but only
%from the anaytic coupling (\ref{basic}), which does contain power
%corrections.
%Let us remark that even if our model doesn't contain free
%parameters to be fitted to the data, in a certain sense we have
%``fitted'' the model itself to data. In fact we have constructed
%our model among different possibilities (f.i. different possible
%prescriptions for the low energy behavior of the QCD coupling)
%with the goal of describing at the best the very accurate data in
%$B$ fragmentation. Then, using the discussed perturbative analogy
%in the resummation formulas,  we have made
% predictions for the spectra $B$ decays.
%
%To summarize, we can say that we include non perturbative power
%corrections of the types $(\Lambda/Q)^p$ with $p\geq 2$
%%(as %predicted by the OPE)
%but  instead of fixing the numerical
%coefficients with an ansatz for the profile of the shape-function
%and fitting the $B$ decays data, we generate power correction with
%an ansatz for the low energy QCD coupling (which reproduce with
%good accuracy both fragmentation and decays data)

Factorization and resummation of threshold logarithms in
semileptonic decays leads to an expression for the
the triple--differential distribution, the most general distribution,
of the following form~\cite{ugo2001}:
%\vspace*{-0.05cm}
\beq
\label{tripla}
\frac{1}{\Gamma} \frac{d^3\Gamma}{dx dw du} \, = \,
C[x, w; \alpha_S(Q)] \, \sigma[u; Q] \, + \,
D[x,u,w; \alpha_S(Q)] \, ,
\eeq
where
%\vspace*{-0.1cm}
%\beq
$x  =  \frac{2 E_l}{m_b}, ~
w =  \frac{Q}{m_b}, ~
u  =  \frac{1 - \sqrt{1 - (2m_X/Q)^2} }{1 + \sqrt{1 - (2m_X/Q)^2} }$
%\eeq
with $E_l$, $E_X$ and $m_X$ being the charged lepton energy, the
total hadron energy and the hadron mass, respectively and the hard
scale is given by $Q = 2 E_X$. $\Gamma=\Gamma(\alpha_S)$ is the
inclusive width of decay~(\ref{start}).
%Furthermore:
%\begin{itemize}
%\item
Furthermore $C[x,w; \alpha_S]$ is a short--distance, process dependent hard factor and
%\item
%$\sigma[u; Q]$ is the universal QCD form factor for heavy--to--light transitions,
%resumming to any order in $\alpha_S$ the series of logarithmically enhanced terms
%to some logarithmic accuracy;
%\item
$D[x,u,w;\alpha_S]$ is a short--distance, process dependent, remainder function,
vanishing in the threshold region $u \to 0$.
% and in lowest--order in $\alpha_S$.
%\end{itemize}
The universal QCD form factor for heavy--to--light transitions,
resumming to any order in $\alpha_S$ the series of logarithmically
enhanced terms to some logarithmic accuracy,
%The heavy flavor decay form factor
has the
following exponential form in the Mellin moments
$N$--space~\cite{cattren}\,\footnote{The Mellin transform of $\sigma(u,Q)$ is as usual
$
\sigma_N(Q) =  \int_0^1 (1-u)^{N-1}  \sigma(u; Q)\, du .$
}$\!\!~^,$\footnote{The QCD form factor $\sigma[u; Q]$ has been numerically computed
for different values of $\alpha_S(m_Z)$ in~\cite{model}.}:
%\vspace*{-0.05cm}
\beq
\label{expoGN}
\log \sigma_N(Q) = \int_0^1 \frac{dy}{y} [ (1-y)^{N-1}\! - 1 ]
\Bigg\{
\int_{Q^2 y^2}^{Q^2 y} \!\frac{dk_{\perp}^2}{k_{\perp}^2}
\tilde{A}[ \tilde{\alpha}_S (k_{\perp}^2)]
+ \tilde{B}[\tilde{\alpha}_S(Q^2 y)]
+ \tilde{D}[\tilde{\alpha}_S(Q^2 y^2)]
\Bigg\}\,,
\eeq
where the functions $A(\alpha_S), ~ B(\alpha_S)$ and
$D(\alpha_S)$ have a standard fixed order expansions in ${\alpha}_S$.
The prescription of our model is simply to replace the standard  functions
$A(\alpha_S), ~ B(\alpha_S)$ and
$D(\alpha_S)$  with the functions
$\tilde{A}(\tilde{\alpha}_S), ~ \tilde{B}(\tilde{\alpha}_S)$ and
$\tilde{D}(\tilde{\alpha}_S)$
obtained from the standard ones by means of the change
of renormalization scheme for the coupling
constant $\alpha_S \to \tilde{\alpha}_S$.

Let us remark that even if our model doesn't contain free
parameters to be fitted to the data, in a certain sense we have
``fitted'' the model itself to data. In fact we have constructed
our model among different possibilities (f.i. different possible
prescriptions for the low energy behavior of the QCD coupling)
(see f.i \cite{ugogiu2004}). A further goal has been to describe
at the best the very accurate data in $B$
fragmentation~\cite{acf}. Then, by using the discussed
perturbative analogy in the resummation formulas,  we have made
 predictions for the spectra $B$ decays~\cite{model}.

Threshold suppression --- in our opinion the main theoretical ingredient
for the measure of $|V_{ub}|$, as discussed in the introduction ---
is represented by the factor
\bea
\label{defW}
W(a,b) &\equiv& \frac{
\Gamma\left[ B \to X_u \ell \nu_{\ell} , p \in (a, b)
\right]} {\Gamma\left[B \, \to \, X_u \, \ell \, \nu_{\ell}\right]}
%\nonumber\\
= \int_{a<p(x,w,u)<b} \frac{1}{\Gamma} \frac{d^3\Gamma}{dx dw
du}dx dw du \, \le \, 1 \, . \eea Note that: \beq
W(p_{\min},p_{\max}) \, = \, \int_{all} \frac{1}{\Gamma}
\frac{d^3\Gamma}{dx dw du}dx dw du \, = \, 1 \, , \eeq where by
``all'' we mean the whole phase--space of $x, \, w$ and $u$. For
discussions and results on threshold resummed spectra of $B
\rightarrow X_u \, \ell\, \nu_{\ell}$ decays at next--to--leading
order see \cite{noi}.

\section{$ |V_{ub}| $ extraction: method}
\label{meth}

%In general, given a kinematical variable $p$,
%such as for example the energy \el\ of the charged lepton, one
%experimentally measures the number of \B's decaying semileptonically to $X_u$
%with $p$ in some interval $(a,b)$, divided by the total number of produced
%\B's (decaying into any possible final state).
The branching ratio  in eq.(\ref{br}) can be computed
theoretically as: \bea {\cal B}\left[ p \in(a, b) \right] \!\!&=&\!\!
\frac{\Gamma\left[ B \to X_u  \ell \, \nu_{\ell} , ~ p
\in (a,b) \right]}{\Gamma[B \to {\rm (anything)}]}
%\nonumber\\
%&=&
=
 \tau_B \, \Gamma \big[ B \, \to \, X_u \, \ell \, \nu_{\ell} \big]
 W(a,b) .
\label{Br1}
\eea
%where
%\beq
%\tau_B \, = \, \frac{1}{ \Gamma\left[ B \to {\rm (anything)}
%\right] } \, .
%\label{Br1}
%\eeq
Since we do not aim at checking QCD but only to extract \Vub , we
can use the experimental measure of the $B$ lifetime, which is
rather accurate\footnote{ At the level of accuracy we are
interested in, we can safely ignore the difference in lifetimes of
the neutral and the charged $B$ mesons. One can take for example
an average life--time. Similarly, we will take average values
between charged and neutral \B's for the other quantities involved
in the rest of the paper. }: \beq \tau_B \, = \, ( \, 1.584 \, \pm
\, 0.007 \, ) \times 10^{-12} \, {\rm s} \, , \eeq and limit
ourself to compute the suppression factor $W(a,b)$ defined in the
previous section (see eq.~(\ref{defW})) and the inclusive
\btoxlnu\ rate: \beq \label{aggiunta} \Gamma\left[ B \, \to \, X_u
\, \ell \, \nu_{\ell} \right] \, = \, \frac{G_F^2 \, m_b^5 \,
|V_{ub}|^2 }{192 \pi^3} \, F(\alpha_S) \, . \eeq QCD corrections
are factorized in the function $F(\alpha_S)$, which is of the
form: \beq F(\alpha_S) \, = \, 1 \, + \, \sum_{n=1}^{\infty} F_n
\, \alpha_S^n \, , \eeq with $F_n \sim \mathcal{O}(1)$ numerical
coefficients\footnote{We can safely take $m_u = 0$.}. A  measure
of \Vub\ is provided by direct comparison of eq.~(\ref{Br1}) with
the corresponding experimental branching ratio. However, within
this method, a large uncertainty comes from the dependence on the
fifth power of the \b--quark
mass in eq.~(\ref{aggiunta})%
%By changing \beq m_b \, \to
%\,m_b \, + \, \delta m \, , \eeq one produces a relative change
%$\epsilon$ of the rate of the order \beq \epsilon \, \approx \, 5
%\, \frac{\delta m}{m_b} \, . \eeq
\,\footnote{For example, a tiny uncertainty
of $\pm 2 \, \%$ on the $m_b$ mass, which is compatible with the
actual estimates, corresponds to about a $\pm 10 \, \%$
uncertainty in the total semileptonic $b \rightarrow u$ width,
translating into a $\pm 5 \, \%$ uncertainty on \Vub.}.

One can eliminate the above (undesired) dependence on $m_b^5$, and
the related uncertainty, by expressing the branching ratio as
follows: \beq \label{baseq} {\cal B}\left[ p \in (a, b) \right] \,
= \, \frac{ {\cal B}_{SL} }{1 \, + \, {\cal R}_{c/u} } \, W(a,b)
\, , \eeq where we have defined the semileptonic branching ratio
($\ell$ is a fixed lepton species; in practice $\ell=e,\mu$): \beq
{\cal B}_{\rm SL} \, \equiv \, \frac{\Gamma(B \, \to \, X_c \,
\ell \, \nu_{\ell}) \, + \, \Gamma(B \, \to \, X_u \, \ell \,
\nu_{\ell})}{\Gamma\left[B \, \to \, {\rm (anything)} \right] }
\eeq and the ratio of $(b \to c)/(b \to u)$ semileptonic widths:
\beq \label{defH} {\cal R}_{c/u} \, \equiv \, \frac{ \Gamma( B \,
\to \, X_c \, \ell \, \nu_{\ell} ) }{ \Gamma( B \, \to \, X_u \,
\ell \, \nu_{\ell} ) } \, . \eeq Since ${\cal B}_{\rm SL}$ is
rather well measured, one can give up on its theoretical
calculation and replace it with the experimental determination
\cite{PDG}: \beq {\cal B}_{\rm SL} \, = \, 0.1066 \, \pm \, 0.0020
\, . \eeq As discussed in the introduction, $W(a,b)$ strongly
depends on the modelling of the threshold region and is calculated
according to the model whose main properties have been summarized
in sec.~\ref{Threshold--Resummation}, for the kinematical
distributions listed in sec.~\ref{res}.

This method presents no $m_b^5$ dependence, since one has to
compute only the ratio of widths ${\cal R}_{c/u} $ and not the
absolute widths. The semileptonic $b \to c$ width is conveniently
written as: \beq \Gamma( B \, \to \, X_c \, \ell \, \nu_{\ell} )
\, = \, \frac{G_F^2 \, m_b^5 \, |V_{cb}|^2 }{192 \pi^3} \, I(\rho)
\, F(\alpha_S) \, G(\alpha_S,\rho) \, , \eeq where \beq \rho \,
\equiv \, \frac{m_c^2}{m_b^2} \, \approx 0.1 \, . \eeq The
function $I(\rho)$ accounts for the suppression of phase--space
because of $m_c \ne 0$ \cite{nir} : \beq I(\rho) \, = \, 1 - 8
\rho + 12 \rho^2 \log \frac{1}{\rho} + 8 \rho^3 - \rho^4 \, . \eeq
Note that there is an (accidental) strong dependence on the charm
mass $m_c$, because of the appearance of a large factor in the
leading term in $\rho$, namely $- \, 8$. As far as inclusive
quantities are concerned, the largest source of theoretical error
comes indeed from the uncertainty in $\rho$. Most of the
dependence is actually on the difference $m_b - m_c$, which can be
estimated quite reasonably within the Heavy Quark Effective Theory
(HQET) --- see next section. Finally, the factor
$G(\alpha_S,\rho)$ contains corrections suppressed by powers of
$\alpha_S$ as well by powers of $\rho$: \beq G(\alpha_S,\rho) \, =
\, 1 \, + \, \sum_{n=1}^{\infty} G_n(\rho) \, \alpha_S^n \, , \eeq
with $G_n(0) = 0$. Note that $ G(0,\rho)  =  G(\alpha_S,0)  =  1
\,$. By inserting the above expressions for the semileptonic
rates, one obtains for the perturbative expansion of ${\cal
R}_{c/u}$: \beq {\cal R}_{c/u} \, = \, {\cal R}_{c/u} \left( \rho,
\alpha_S, |V_{ub}|/|V_{cb}| \right) \, = \,
\frac{|V_{cb}|^2}{|V_{ub}|^2} \, I(\rho) \, G(\alpha_S, \rho) \, .
\eeq Let us stress that this method actually provides a measure of
the ratio $|V_{ub}|/|V_{cb}|$, but since the error on $|V_{cb}|$
is rather small and theoretically well understood, one is
basically measuring \Vub.\footnote{ The average of
determinations of \Vcb\ coming from a global fit to the \btoxclnu\ and
$b\rightarrow s \gamma$ moments in the kinetic and 1S schemes,
in good agreement with each other,
is \Vcb $= ( 41.6 \pm 0.6) \times 10^{-3} $ \cite{hfag,PDG}. }

We prefer to use the method
%of measuring \Vub\ involving
based on the computation of the ratio of $(b\to c)/(b \to u)$
semileptonic rates ${\cal R}_{c/u}$ (see eq.~(\ref{defH})),
instead of the method involving the {\it absolute} $b \, \to \, u
\, \ell \, \nu_{\ell}$ rate (see eq.~(\ref{aggiunta})), because in
the former case only ratios of inclusive widths need to be
evaluated. As a consequence, while eq.~(\ref{aggiunta}) depends on
the absolute value of \mb, eq.~(\ref{defH}) depends mostly on the
difference \mb$-$\mc,
%in which many hadronic uncertainties %possibly cancel.
opening the possibility of partial cancellation of hadronic
uncertainties. However, the two methods  give compatible values of
\Vub\ and the method based on eq.~(\ref{aggiunta}) is used to
estimate the systematic error on the method based on
eq.~(\ref{defH}). The fact that the two methods give similar
results corroborates the conclusion that differences with respect
to previous analyses originate from a smaller suppression in the
threshold region in our model compared to previous models, rather
than from different estimates of inclusive quantities, i.e.
different choices of quark masses, QCD coupling, etc. In other
words, we do not obtain a small value of \Vub\ by an ad--hoc
choice of quark masses \mb\ and \mc\ or of $\alpha_S(m_Z)$. The
difference lies in the different modelling of the threshold
region.

\subsection{Quark masses}
\label{massebec}

Since quarks are confined inside observable hadrons, their masses cannot be directly
measured and their values are biased by the selected theoretical
framework. It is well known that pole masses are affected by
the poor behavior of the perturbative series relating them to
physical quantities; replacing the pole mass with the
$\overline{\mbox{MS}}$ mass in dimensional regularization
slightly improves the asymptotic behavior of the series. Several
alternative definitions of the \b--quark mass have been introduced
in the literature~\cite{masses}, in order to give a better convergence
of the first (few) orders of the perturbative series and
consequently reduce the theoretical errors.

We have performed the calculation  in the $\overline{\mbox{MS}}$
mass scheme. The $\overline{\mbox{MS}}$ masses for the $b$ and the
$c$ quark are taken $\overline{m}_b(\overline{m}_b)= 4.20 \, \pm
0.07 \, \rm{GeV}$ and $ \overline{m}_c(\overline{m}_c)= 1.25 \,
\pm 0.09 \, \rm{GeV}$ \cite{PDG}, respectively. However, in order
to take into account the uncertainties coming from a different
scheme definition, we have considered the pole--mass scheme as
well. The HQET in lowest order (static theory) gives the following
relation for the difference of the on--shell beauty and charm
mass: \beq \label{HQET2_mass} m_b \, - \, m_c \, = \,  m_B \, - \,
m_D \, + \, \mathcal{O} \left[ \Lambda^2 \left( \frac{1}{m_c} -
\frac{1}{m_b} \right) \right] \eeq with \beq m_B \, - \, m_D \,
\simeq \, 3.41 ~\gev \, . \eeq

This value  is also consistent with the mass estimates in the
$\rm{kin_{EXP}}$  and $1 \,S_{\rm{EXP}}$ schemes \cite{bauer}. The
leading corrections to (\ref{HQET2_mass}),
$\mathcal{O}[\Lambda^2(1/m_c-1/m_b)]$, involve the chromomagnetic
operator and the kinetic one. One can cancel the chromomagnetic
corrections by taking the ``spin--averaged" meson
masses~\cite{ugo-mass}: \beq \label{HQET_mass} m_b \, - \, m_c
\simeq \frac{m_B + 3 \,m_{B^{\star 0}}}{4} \, - \, \frac{m_D +
3\,m_{D^{\star 0}}}{4} \, + \, \cdots \, , \eeq where the smaller
difference of $\approx 70$ MeV, implies, for a fixed beauty mass,
a heavier charm quark mass\footnote{We use $m_{\Bz} = 5279.50 \,
\pm 0.33 \, \mev$, $m_{B^{\star 0}} = 5325.1 \, \pm 0.5 \, \mev$,
$m_{\Dz} = 1864.84 \, \pm 0.17 \, \mev$ and $m_{D^{\star 0}}=
2006.97 \, \pm 0.19 \, \mev$ \cite{PDG}.}: \beq \frac{m_B + 3
\,m_{B^{\star 0}}}{4} \, - \, \frac{m_D + 3\,m_{D^{\star 0}}}{4}
\, \simeq \, 3.34 ~ \gev \, . \eeq For a consistent subleading
computation one also needs the kinetic contribution, which cannot
be extracted from the data and it is computed
theoretically~\cite{bauer,Hoang}.
%One then obtains a band which
%contains the static theory prediction
%(\ref{HQET2_mass})\
The conclusion is that we consider safe to eliminate the $c$ pole
charm mass by using a range of $m_b-m_c$ given by
eqs.~(\ref{HQET2_mass}) and (\ref{HQET_mass})\footnote{In order to
check eqs.~(\ref{HQET2_mass}) and (\ref{HQET_mass}), we have
related $m_b$ and $m_c$ by considering the ratio between $B\to X_c
\,\tau\, \nu$ and $B\to X_c\, e\, \nu$ branching ratios, which
have been measured with a reasonable accuracy at LEP and
SLD~\cite{PDG}.
%\footnote{We neglect $B\to X_u\, \tau\, \nu_\tau$
%decays which constitute at most a $10\%$ of $B\to X\, \tau\, \nu_\tau$}:
%\begin{equation}
%\label{exp}
%X_{\rm exp} \, = \, \frac{ { \cal B}(B \, \to \, X_c \, \tau \, \nu_\tau) }
%{ { \cal B}(B \, \to \, X_c \, e \, \nu_e) }
%\, = \, \frac{ ( 2.41 \, \pm \, 0.23 ) \, \%}{ ( 10.86 \, \pm \, 0.35 ) \, \% } \, .
%\end{equation}
The corresponding theoretical quantity strongly depends on $m_b$
and $m_c$~\cite{czarn}:
%\begin{equation}
%{X}_{\rm th}( m_b, \, m_c, \, \alpha_S ) \, \equiv \, \frac{\Gamma(
%b \, \to \, c \, \tau \, \nu_\tau )} {\Gamma( b \, \to \, c \, e
%\, \nu_e )}. \label{th1}
%\end{equation}
%By equating eqs.~(\ref{exp}) and (\ref{th1}), one obtains an almost
% horizontal strip in
%the $m_c$ versus $m_b$ plane; the central value is practically coincident
%with the line given by the difference $m_b - m_c$  in eq.~(\ref{HQET2_mass}),
%but the band also covers the line related to the improved relation
%(\ref{HQET_mass}).
%the
the $b$ and $c$ masses extracted in this way %$X$ therefore
corroborates the quark mass relations of the effective theory.}.
%The static theory value (\ref{HQET2_mass}) is also consistent with the estimates in the
%$\rm{kin_{EXP}}$  and $1 \,S_{\rm{EXP}}$ schemes
%\cite{bauer}:
%\begin{eqnarray}
%m_b - m_c &=& 3.41 \pm 0.01 \, \rm{GeV} \qquad (1\, S_{\rm{EXP}})
%\nonumber \\
%m_b - m_c &=& 3.40 \pm 0.01 \pm 0.01 \,\rm{GeV} \qquad
%\rm{(kin_{EXP})}
%\end{eqnarray}
%The second error for the $\rm{kin_{EXP}}$ scheme is the shift due
%to changing $\mu$ from $1$ to $1.5$~\gev..

\section{Results}
\label{res}

This is the central section of the paper in which we determine
\Vub\ from measured semileptonic branching fractions, in limited
regions of the phase--space, and we perform the corresponding
averages. The experimental analyses are categorized according to
the kinematical distribution looked at, where selection criteria
are applied to define the limited phase--space on which the
branching ratio is computed. The analyses are:

\begin{enumerate}
\item \el: where the distribution looked at is the lepton energy
(\el). There are results from \babar~\cite{babar-el},
Belle~\cite{belle-el} and CLEO~\cite{cleo-el}. The lepton energy
ranges from 1.9, 2.0 and 2.1~\gev\ for Belle, \babar, and CLEO,
respectively, up to 2.6~\gev. As described in ref.~\cite{model},
we will look only at the range where data are not affected by
potential $b\to c\, \ell\, \nu_\ell$ background: 2.3~\gev $<$ \el
$<$ 2.6~\gev . \item \mx : where the distribution looked at is the
invariant mass of the hadron final state (\mx). Both
\babar~\cite{babar-breco} and Belle~\cite{belle-breco} have
performed one analysis for \mx $<$ 1.55~\gev\ and \mx $<$ 1.7~\gev
, respectively. The selection strategy is based on the full
reconstruction of the $B$ meson on the other side of the event
(\B--reconstruction analysis). The same selected events are looked
at also by using two other kinematical distributions: \mxqsq\ and
\pplus, as we will see below. In all the three cases a lower cut
on the photon energy at 1~\gev\ is applied.
\item \pplus: where the distribution
looked at is $ \pplus \equiv E_X - |\vec{p}_X|$, $E_X$ and
$\vec{p}_X$ being the energy and the magnitude of the 3--momentum
of the hadronic system, respectively. The \B--reconstruction
analysis is used. There are both \babar~\cite{babar-breco} and
Belle~\cite{belle-breco} results. Both analyses require $P_+ <$
0.66~\gev. \item \mxqsq : where the distribution looked at is a
two dimensional distribution in the plane of the hadronic mass and
the transferred squared momentum $q^2$ to the lepton pair. The
analyses from \babar\ and Belle are described in
\cite{babar-breco} and \cite{belle-breco}, respectively, using the
\B--reconstruction analysis. Moreover, another technique, called
simulated annealing, is used by Belle to select events in the
\mxqsq\ plane~\cite{belle-ann}. Events with hadron mass lower than
1.7~\gev\ and $q^2$ larger than 8~$\gev^2$ are selected in all the
cases. \item \elsmax : where the distribution looked at is a two
dimensional distribution in the electron energy and \smax,  the
maximal $\mx^2$ at fixed $q^2$ and \el. There is a result from
\babar~\cite{babar-elsmax}. The requests on the kinematic
variables are \el $> 2.0\gev$ and $\smax\ < 3.5\gev^2$.
\end{enumerate}

We compute \Vub\ for each of the analyses starting from the
corresponding partial branching fractions. Then, we determine
the average \Vub\ value using the HFAG methodology~\cite{hfag},
together with Ref.\cite{rogerbarlow}.

%Averaged values of \Vub\ are presented in the case of all the
%uncorrelated measurements, where the uncorrelation refers to the
%experimental errors, and in the case of each category of kinematic variables.
%There are large correlations among the three kinematical
%distributions in the \B--reconstruction analysis, which are not
%reported in the experimental papers. Thus, we use only the \mx\
%analysis, which is the one with the smallest experimental error,
%when performing the average on all the available measurements.
Table~\ref{tab:global_average} reports the extracted values of
\Vub\ for all the analysis methods and their corresponding
average. The errors are experimental (i.e. statistical and
systematic) and theoretical, respectively.
The average is:
\beq
\Vub \, = \, ( \, 3.76 \, \pm \, 0.13 \, \pm \, 0.22  \, ) \times 10^{-3} \, ,
\eeq
consistent with the measured value of \Vub\ from
exclusive decays~\cite{hfag} and the indirect measure~\cite{utfit}.
The correlation among the analyses has been taken into
account when performing the average~\footnote{Concerning the \B--reconstruction analysis, we use only the \mx\
analysis in Table~\ref{tab:global_average}, according to the approach in
Ref.~\cite{hfag}.}.

\begin{figure}[!htb]
\vspace{0.3in}
 \begin{centering}
  \includegraphics[width=0.54\textwidth,totalheight=10.8cm]{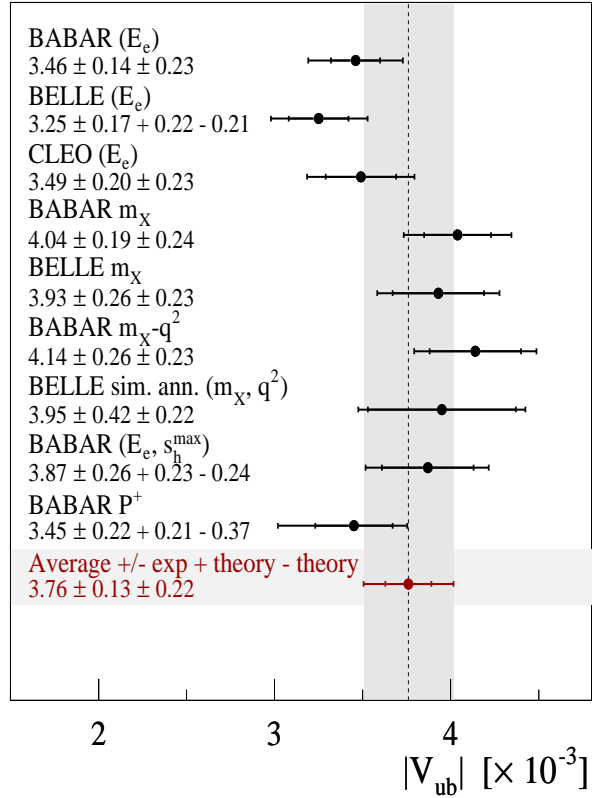}
  \caption{\it \Vub\ values for the analyses used in the average and their average.}
  \label{fig:vub}
 \end{centering}
 \end{figure}

The table shows also the
criteria used for the determination of the partial branching ratio
($\Delta{\cal B}$).
%The theoretical errors are considered
%completely correlated among all the experimental analyses, when
%performing the average.
The \Vub\ values and the corresponding average are plotted in
Figure~\ref{fig:vub}.

\begin{table*}[!htb]
\begin{center}
\caption{\it The first column in the table shows the
analyses used in the average, the second column shows the corresponding values of
\Vub, and finally the last column shows the criteria for which
$\Delta{\cal B}$ is available. The final row shows the average
value of \Vub. The errors on the \Vub\ values are experimental and
theoretical, respectively. The experimental error includes both
the statistical and systematic errors.}
\vspace{0.1in}
\begin{tabular}{l|c|c}
\hline\hline
Analysis&   \Vub ($10^{-3}$)           &  $\Delta{\cal B}$ criteria \\
\hline
\babar\ (\el)  \cite{babar-el}  &   3.46$\pm$ 0.14 $^{+0.23}_{-0.23}$  &  \el $ >2.3$\gev\\
Belle (\el)  \cite{belle-el}  &   3.25$\pm$ 0.17 $^{+0.22}_{-0.21}$  &  \el $ >2.3$\gev\\
CLEO  (\el)  \cite{cleo-el}   &   3.49$\pm$ 0.20 $^{+0.23}_{-0.23}$  &  \el $ >2.3$\gev\\
\babar\ (\mx)   \cite{babar-breco}   &   4.04$\pm$ 0.19 $^{+0.24}_{-0.24}$  & \mx $<1.55$~\gev\\
Belle (\mx)   \cite{belle-breco}   &   3.93$\pm$ 0.26 $^{+0.23}_{-0.23}$  & \mx $<1.7$~\gev\\
\babar\ (\mxqsq)   \cite{babar-breco}   &   4.14$\pm$ 0.26 $^{+0.23}_{-0.23}$  & \mx $<1.7$~\gev, $q^2>8$~$\gev^2$ \\
Belle \mxqsq   \cite{belle-ann}   &   3.95$\pm$ 0.42 $^{+0.22}_{-0.22}$  &\mx $<1.7$~\gev, $q^2>8$~$\gev^2$ \\
\babar\ \elsmax \cite{babar-elsmax}&   3.87$\pm$ 0.26 $^{+0.23}_{-0.24}$  & \el $ >2.0$~\gev , \smax$<3.5$~$\gev^2$\\
\babar\ (\pplus)   \cite{babar-breco}   &   3.45$\pm$ 0.22 $^{+0.21}_{-0.37}$  & \pplus $<0.66$~\gev\\
\hline
Average                            &   3.76$\pm$ 0.13 $^{+0.22}_{-0.22}$  &\\
\hline\hline
\end{tabular}
\label{tab:global_average}
\end{center}
\end{table*}
\vspace{0.1in}

Several sources of theoretical errors have been considered:
\begin{itemize}
\item
in addition to our preferred method based on eq.~(\ref{baseq}),
we also use the method based on eq.~(\ref{aggiunta}) to extract
the value of \Vub.
Since the two methods described in the previous section
basically involve different inclusive quantities,
this error allows a cross--check of their evaluations,
i.e. basically of the choices of the $b$ and $c$ masses adopted;

\item
we compute inclusive quantities both in the $\overline{\mbox{MS}}$
and pole schemes for the quark masses.
The \Vub\ value using the $\overline{\mbox{MS}}$ masses is our default,
the value using the pole scheme masses, where the ranges defined
in sec.~\ref{massebec} are used, gives the systematic uncertainty.
Since in general higher--order corrections are
different in the two schemes, that should provide an estimate of the size
of unknown higher--order effects.

\item
we vary the order at which the rate is computed from the exact NLO to the
approximate NNLO~\cite{rit-cza}.
Since the perturbative series for QCD is believed to
be an asymptotic one and $\alpha_S = \alpha_S(m_b) \approx 0.22$ in $b$ physics
is rather large, that should provide a reasonable estimate on the truncation error;

\item
we vary all the parameters which enter in
the computation of \Vub\ within their errors, as given by the
PDG~\cite{PDG}.
\end{itemize}
What we {\it cannot change} is the modelling of the threshold
region represented by the factor $W(a,b)$, which is fixed in our
model, as discussed in the introduction. How good it such
modelling of the threshold region can only be estimated
indirectly, by considering different decay spectra, where
presumably threshold effects enter in different ways.

Future work towards improving the determination of
the \b--quark mass in the $\overline{\mbox{MS}}$ and the pole scheme will help in reducing
our error on \Vub .

Table~\ref{tab:theo_errors} reports the fractional contributions
to the theoretical errors due to the different sources.
A large excursion of the error among the analyses is due to
$\alpha_S$, which varies from a minimum of $\pm 0.6\%$ for the \mxqsq\
analyses to a maximum of $\pm 3.5\%$ for the endpoint analyses.
The largest contributions to the error are given by the charm mass
in the $\overline{\mbox{MS}}$ scheme ($\pm 4.4\%$) and the variation from the
$\overline{\mbox{MS}}$ to the pole scheme (from $-\,1.3\%$ to $-\,5.2\%$
starting from lower to higher masses),
due to the conservatively larger error used for the pole scheme mass.

Table~\ref{tab:partial_averages} shows the \Vub\ averages for
different analysis categories. Note that the \mxqsq\ analyses tend
to have the largest values of \Vub, while the endpoint analyses
the smallest.

\begin{table*}[!htb]
\vspace{-.1in}
\begin{center}
\caption{\it The first column of the table shows the different
contributions to the theoretical errors, the second column shows
the corresponding variation, and finally the third column shows
the percentage contribution with respect to the \Vub\ value.}
\vspace{0.1in}
\begin{tabular}{l|c|c}
\hline\hline \multicolumn{3}{c}{Theoretical Errors}
\\\hline
Contribution & Variation & Error (\%)
\\\hline
$\alpha_S$ & $0.1176 \pm 0.0024$ & $\pm 0.6 \rightarrow 3.5$
\\
\Vcb &$(41.6 \pm 0.6)\times 10^{-3}$ & $\pm 1.4$
\\
\mb\ (\gev) &$4.20 \pm 0.07 $ & $\pm 0.6$
\\
\mc\ (\gev) &$1.25 \pm 0.09 $ & $\pm 4.4$
\\
${\cal B}$ (\btoxlnu)  &$0.1066 \pm 0.0020$& $\pm 1.0$
\\
\Vub\ method && $+0.8$
\\
pole mass (\gev)&
$\begin{array}{c}
4.7<m_b<5.0,\,\,1.47<m_c<1.83\\
3.34<m_b-m_c<3.41
\end{array}$
&$-1.3 \rightarrow  -5.2$
\\
approx. NNLO rate && $+2.0$
\\
\hline\hline
\end{tabular}
\label{tab:theo_errors}
\end{center}
\end{table*}

\begin{table*}[!htb]
\begin{center}
\vspace{-.1in}
\caption{\it The table contains the \Vub\ values for several analyses
and the corresponding averages. The errors on the \Vub\ values are
experimental and theoretical, respectively. The experimental error
includes both the statistical and systematic errors.}
\vspace{0.1in}
\begin{tabular}{l|c|c}
\hline\hline \multicolumn{3}{c}{\Vub\ for endpoint analyses ($10^{-3}$)}\\\hline
\babar\ (\el) \cite{babar-el}    &   3.46$\pm$ 0.14 $^{+0.23}_{-0.23}$& \el $>2.3$ \gev \\
Belle (\el) \cite{belle-el}      &   3.25$\pm$ 0.17 $^{+0.22}_{-0.21}$ & \el $>2.3$ \gev\\
CLEO (\el) \cite{cleo-el}        &   3.49$\pm$ 0.20 $^{+0.23}_{-0.23}$ &  \el $>2.3$ \gev\\
\hline Average                   &   3.43$\pm$ 0.15 $^{+0.23}_{-0.22}$  &\\
\hline\hline \multicolumn{3}{c}{\Vub\ for \mx\ analyses ($10^{-3}$)}\\\hline
\babar\ (\mx)   \cite{babar-breco}    &   4.04$\pm$ 0.19 $^{+0.24}_{-0.24}$  &\mx $<1.55$~\gev \\
Belle (\mx)   \cite{belle-breco}      &   3.93$\pm$ 0.26 $^{+0.23}_{-0.22}$ &\mx $<1.7$~\gev \\
\hline Average                        &   4.00$\pm$ 0.16 $^{+0.24}_{-0.23}$ &\\
\hline\hline \multicolumn{3}{c}{\Vub\ for \mxqsq\
analyses ($10^{-3}$)}\\\hline
\babar\ \mxqsq   \cite{babar-breco}  &   4.14$\pm$ 0.26 $^{+0.23}_{-0.23}$ &\mx $<1.7$~\gev, $q^2>8~\gev^2$\\
Belle \mxqsq   \cite{belle-breco}    &   4.21$\pm$ 0.37 $^{+0.23}_{-0.23}$ &\mx $<1.7$~\gev, $q^2>8~\gev^2$\\
Belle \mxqsq   \cite{belle-ann}      &   3.95$\pm$ 0.42 $^{+0.22}_{-0.22}$ &\mx $<1.7$~\gev, $q^2>8~\gev^2$\\
\hline Average                    &   4.13$\pm$ 0.21 $^{+0.23}_{-0.23}$ &\\
\hline\hline \multicolumn{3}{c}{\Vub\ for \pplus\
analyses ($10^{-3}$)}\\\hline
\babar\ (\pplus) \cite{babar-breco}  &   3.45$\pm$ 0.22 $^{+0.21}_{-0.37}$ &$P^+ <0.66$ \\
Belle (\pplus) \cite{belle-breco}    &   3.73$\pm$ 0.32 $^{+0.23}_{-0.29}$&$P^+ <0.66$ \\
\hline Average                       &   3.55$\pm$ 0.19 $^{+0.21}_{-0.23}$ &\\
\hline\hline
\end{tabular}
\label{tab:partial_averages}
\end{center}
\end{table*}

The larger value of \Vub coming from the analysis of the double
distribution in $(m_X, q^2)$ is expected on qualitative basis. For
$m_X = 0$, the hard scale $Q$ is given by \beq Q \, = \, m_\B \, -
\, \frac{q^2}{m_B} \, < \, 0.71 \, m_\B \, , \eeq because $q^2 \,
> \, 8 ~ \gev^2$. The lower cut on $q^2$ therefore significantly
reduces the hard scale $Q$ from the ``natural'' value $Q = m_\B$.
The point is that our model has been constructed to describe
\B--decay spectra having the (maximal) hard scale $Q = m_\B$, and
not spectra having a smaller hard scale. Indeed, the model was
checked against beauty fragmentation at the $Z^0$ peak~\cite{acf},
where the dominant infrared effects are controlled by a hard scale
equal to $m_\B$. In other words, to analyze the $(m_X, q^2)$
distribution, we are using the model in a region where it has not
been checked and it is no surprise that it does not work so well
in this case.

\section{Conclusions}
\label{concl}

We have analyzed semileptonic $B$ decay data in the framework of a
model for QCD non--perturbative effects based on an effective
time-like QCD coupling, free from Landau singularities.
The analysis has considered the
kinematical distributions in \el, \mx, and \pplus,  as well as the
two dimensional distributions in \mxqsq\ and \elsmax, taking into
account the experimental kinematical cuts.

Our inclusive measure of the \Vub CKM matrix element is: \beq \Vub
\, = \, ( \, 3.76 \, \pm \, 0.13 \, \pm \, 0.22 \, ) \times
10^{-3} \, . \eeq The errors on the \Vub\ value are experimental
and theoretical, respectively. The experimental error includes
both the statistical and systematic errors.

For the first time, an inclusive value for \Vub\ is obtained
which is in complete agreement with the exclusive determination.
%($0.36\,\sigma$).
Current literature presents a  discrepancy among
previous inclusive determinations of \Vub\ on one side and the exclusive
determinations ($\approx 2\,\sigma$)
and the over--all fit of the Standard Model
($\approx 3\,\sigma$)
on the other side\cite{hfag}.

Let us try to identify the differences between our approach and
the previous ones.
A first difference lies in the selected
lepton energy range. According to our
%model,
analysis, lepton spectra below $\approx 2.3$~\gev\ measured at the
\B--factories suffer from an under--subtracted charm background.
Because of that, we have limited our analysis to pretty large
lepton energies $\el > 2.3$~\gev. If we take a smaller cutoff, we
obtain up to a $\approx +2.7\%$ larger average value of $\Vub$, in
order to simulate $b \to c$ events. As far as theory is concerned,
our model seems to produce a smaller Sudakov suppression compared
to other models constructed on top of soft--gluon resummation,
such as for example the dressed gluon exponentiation~\cite{gardi}.
For a fixed experimental rate, a smaller Sudakov suppression
implies indeed larger hadronic form factors and smaller \Vub's. As
discussed above,
%we are rather confident in our model for
%phenomenological reasons rather than theoretical ones:
our confidence in the model is also based on phenomenological
grounds; we have checked it in beauty fragmentation~\cite{acf},
where soft contributions are similar to those in $b$
decay\footnote{ The soft effects are contained in the initial
condition of the $b$ fragmentation function $D^{\rm ini}$, which
has the same resummed expression as the shape function in $b$
decays~\cite{sf}. }.
% Even though our model of soft--gluon dynamics
%is formally without free unknown parameters, we may say we have
%constructed it, out of many possibilities, as a kind of ``expert
%system''. Once ``trained'', by giving beauty fragmentation data in
%input, it should predict reasonable beauty decay spectra.

In conclusion, the inclusive extraction of \Vub requires the
calculation of inclusive quantities, strongly dependent on $b$ and
$c$ masses, as well as the evaluation of threshold--suppressed
quantities, the latter containing large infrared logarithms and
Fermi--motion (non--perturbative) effects. We argue that the main
difference of our model with respect to previous ones is a smaller
suppression of the threshold region.

%\newpage
\vskip 0.5truecm

\centerline{\bf Acknowledgments}

\vskip 0.4truecm

\noindent
This work is supported in part by the EU Contract No.
MRTN-CT-2006-035482 ``FLAVIAnet''.

\end{document}